\documentclass[aps,
               prb,
               reprint,
               groupedaddress,
               showpacs,
               citeautoscript,
               floatfix]{revtex4-1}

\usepackage{url}
\usepackage[breaklinks]{hyperref}
\usepackage{breakurl}

\usepackage{amsmath}
\usepackage{amssymb}
\usepackage{graphicx}
\usepackage{dsfont}
\usepackage{extarrows}

\newcommand{\op}[1]{\hat{#1}}
\newcommand{\Hamilton}{\mathbf{H}}
\newcommand{\SpectralFct}{\mathbf{A}}
\newcommand{\G}{\mathbf{G}}
\newcommand{\GRetarded}{\G^{\textnormal{R}}}
\newcommand{\GAdvanced}{\G^{\textnormal{A}}}
\newcommand{\SelfEnergy}{\mathbf{\Sigma}}
\newcommand{\ContourOrdering}{\mathcal{T}_C}
\newcommand{\up}{\uparrow}
\newcommand{\down}{\downarrow}

\DeclareMathOperator*{\sumint}{%
\mathchoice%
  {\ooalign{$\displaystyle\sum$\cr\hidewidth$\displaystyle\int$\hidewidth\cr}}
  {\ooalign{\raisebox{.14\height}{\scalebox{.7}{$\textstyle\sum$}}\cr\hidewidth$\textstyle\int$\hidewidth\cr}}
  {\ooalign{\raisebox{.2\height}{\scalebox{.6}{$\scriptstyle\sum$}}\cr$\scriptstyle\int$\cr}}
  {\ooalign{\raisebox{.2\height}{\scalebox{.6}{$\scriptstyle\sum$}}\cr$\scriptstyle\int$\cr}}
}

\begin{document}

\title{How to interpret the spectral density of the Keldysh nonequilibrium Green's function}

\author{K.J. Pototzky and E.K.U. Gross} 

\affiliation{Max Planck Institute of Microstructure Physics, 
             06120 Halle (Saale), Germany}

\date{\today}

\begin{abstract}
    This paper is devoted to the 
    study and interpretation of the spectral function $\SpectralFct(\omega, T)$
    of the Keldysh nonequilibrium Green's function.
    The spatial diagonal of the spectral function
    is often interpreted as a time-dependent
    local density of states. We show that 
    this object 
    can take negative values
    implying that
    a simple probability interpretation
    as a time-dependent density of states
    is not possible.
    The same issue also occurs
    for the Wigner function $P(x,p)$ where it 
    is solved by taking the uncertainty principle
    into account. 
    We follow the same path
    and incorporate the 
    time-energy uncertainty relation 
    to define a convoluted spectral function
    that allows for a probability 
    interpretation. The usefulness of this
    quantity as a interpretative tool
    is demonstrated by visualizing 
    the charge dynamics in a 
    quantum dot coupled to 
    superconducting leads.
\end{abstract}


\maketitle

\section{Introduction}
    \label{sec:Introduction}
    The ongoing miniaturization
    of electronics may 
    ultimately lead to the use
    of single 
    molecules as its building blocks.
    A sound theoretical 
    understanding of phenomena in nanojunctions
    is therefore of great importance.
    Since the first proposal
    of using a molecule as an electronic component 
    by Aviram and Ratner in 1974 \cite{Aviram1974},
    an enormous number of research
    articles have appeared. 
    Several textbooks, e.g. Refs \onlinecite{Cuniberti2005, DiVentra2008, Cuevas2010},
    serve as excellent introductions 
    to the field.
    
    While, traditionally, the prime quantity of interest
    was the current-voltage characteristics
    of the molecular junction,
    calculated or measured in the steady
    state, there has been a shift of attention towards
    time-resolved studies of quantum transport
    in recent years \cite{Myoehaenen2009, Myoehaenen2010, Uimonen2010, Uimonen2011, Khosravi2012}.
    With this type of studies one may address questions like:
    How much time does it take until the steady state
    is reached and, by which structural changes
    in the junction, can this switching time be made
    shorter or longer? Is there a steady
    state at all? If there is a steady state, is it unique?
    If it is not unique, how can one switch between multiple steady states?
    On the theoretical side, various approaches
    have been put forth to study the real-time
    dynamics of molecular junctions.
    Among those are the Kadanoff-Baym equations
    \cite{Stefanucci2013,Myoehaenen2009, Myoehaenen2010, Knap2011}
    representing the time-dependent variety of many-body
    perturbation theory,
    time-dependent density functional theory
    \cite{RungeGross1984, Baer2004, Burke2005, Kurth2005, YuenZhou2009, DiVentra2007, Appel2009, Appel2011},
    the time-dependent tight binding approach \cite{Wang2011b,Zhang2013,Oppenlaender2013},
    the hierachy equation of motion approach \cite{Jin2008, Zheng2008, Zheng2008b},
    the multi-configuration time-dependent
    Hartree-Fock method \cite{Zanghellini2004, Wang2009, Wang2013, Albrecht2012}
    as well as Quantum Monte-Carlo \cite{Muehlbacher2008}.
    
    Once the numerical time propagation of the respective
    equation of motion has been performed,
    the next question is about the tools to 
    interpret and visualize the results.
    One possibility is the time-dependent
    electron localization function
    \cite{Burnus2005,Raesaenen2008},
    a correlation function 
    suitable to visualize chemical bonds.
    
    In this article we investigate another
    quantity, the spectral function 
    $\SpectralFct(\omega, T)$ of the
    Keldysh non-equilibrium Green's function.
    This object is more targeted towards
    the understanding of charger-transfer
    processes and has provided valuable insights
    in the internal dynamics of molecular
    junctions \cite{Myoehaenen2009, Myoehaenen2010, Uimonen2010, Uimonen2011, Khosravi2012}.  
    The definition of the spectral functions \cite{Stefanucci2013} is
    \begin{align}
        \label{eqn:Def-SpecFct-Energy-Time}
        \SpectralFct(\omega, T) &= \int_{-\infty}^{\infty}\frac{\,d\tau}{2\pi} e^{i\omega \tau} \SpectralFct\left(T + \frac{\tau}{2}, T - \frac{\tau}{2}\right),\\
        \label{eqn:Def-SpecFct-Time-Time}
        \SpectralFct(t,t')      &= i \left[\G^> (t,t')  - \G^< (t,t')\right]
    \end{align}
    where $\G^\gtrless(t,t')$ are the standard greater (lesser)
    nonequilibrium Green's functions \cite{Stefanucci2013}.
    $\SpectralFct(\omega, T)$ is a matrix with respect to
    some representation referring, e.g., to
    space and spin coordinates $\SpectralFct_{r\sigma,r'\sigma'}(\omega, T)$
    or to localized orbitals $\SpectralFct_{i,j}(\omega, T)$.
    The objective of this article is to give a clear-cut
    physical interpretation of the 
    diagonal of this matrix. In particular
    we shall investigate whether and to which 
    extend it can be viewed as a 
    time-dependent density of states.

    The paper is structured as follows:
    In the next section, we explain 
    the model of a quantum dot coupled
    to superconducting leads,
    state the corresponding
    Hamiltonian and define all
    necessary nonequilibrium Green's functions. 
    In section \ref{sec:calc-spectral-function}
    we derive a
    method to calculate the 
    large-time behaviour 
    of $\SpectralFct(\omega, T)$
    directly from the defining
    equations. 
    We further present a second method
    using single particle wave functions which
    give access to 
    $\SpectralFct(\omega, T)$ 
    at all times $T$.
    This, in particular, allows the 
    visualization of switching effects.
    In section \ref{sec:results-negative-values},
    we show with a simple example that
    the probability 
    interpretation of the spectral function
    is generally not correct
    because it can take negative values.
    We solve this problem 
    by taking the time-energy uncertainty relation
    into account.
    In section \ref{sec:results-switching-processes}, we study
    the spectral function
    in two situations
    with a change in the bias.
    First, we switch on the bias in a step-like fashion
    and look at the spectral function
    decomposed into contributions 
    of scattering and bound states.
    Second, we visualize the 
    spectral function of the Andreev bound states
    under the influence of a rectangular bias pulse.
    The final section \ref{sec:conclusion} summarizes 
    the outcome of the presented work.

\section{Theoretical foundation}
    \label{sec:theoretical-foundation}
    \subsection{The Model}
        
        \begin{figure}[htb]
            \begin{center}
                \includegraphics[scale=1.0]{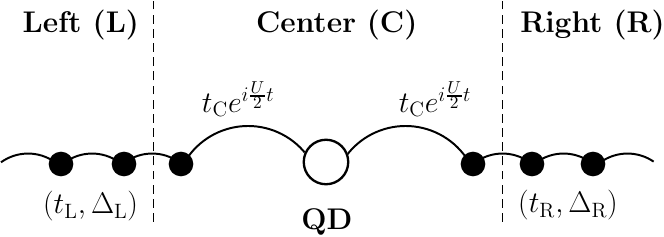}
                \caption{A sketch of the tight binding model.
                         The filled dots represent the sites
                         of the superconducting leads
                         while the central site (the quantum dot)
                         is normal conducting.
                         The left and right site of the central 
                         region are buffer sites of the leads.
                         }
                \label{fig:Sketch-TB-Model}
            \end{center}
        \end{figure}
        
        The model is a tight binding chain of 
        atoms where the central region
        consists of a quantum dot  (QD)
        and two buffer sites of the leads, one
        in each direction. A sketch is
        shown in figure \ref{fig:Sketch-TB-Model}.
        The sites are enumerated from left to right
        with $0$ being the quantum dot (QD).
        Filled dots represent superconducting sites,
        the empty one is normal-conducting.
        
        The bias enters the model Hamiltonian
        using Peierls' substitution \cite{Peierls1933}.
        The Hamiltonian reads as
        \begin{align}
            \op{H}                          &= \op{H}_{\textnormal{Center}}(t) + \op{H}_{\textnormal{tunnelling}} +  \op{H}_\textnormal{Leads}\\
            \op{H}_{\textnormal{Center}}(t) &= \sum_{\sigma \in \{\uparrow, \downarrow\}} \left( t_{\textnormal{C}}e^{i\frac{U}{2}t} \op{c}_{-1,\sigma}^\dagger \op{c}_{\textnormal{QD},\sigma} + H.c. \right)\\
                                            & \quad + \sum_{\sigma \in \{\uparrow, \downarrow\}} \left( t_{\textnormal{C}}e^{i\frac{U}{2}t} \op{c}_{\textnormal{QD},\sigma}^\dagger \op{c}_{1,\sigma} + H.c. \right) \nonumber \\
                                            & \quad + \left(\Delta_{\textnormal{L}} \op{c}_{-1,\uparrow}^\dagger \op{c}_{-1,\downarrow}^\dagger + H.c. \right) \nonumber \\
                                            & \quad + \left(\Delta_{\textnormal{R}} \op{c}_{1,\uparrow}^\dagger \op{c}_{1,\downarrow}^\dagger + H.c. \right), \nonumber \\
            \op{H}_{\textnormal{tunnelling}}&= \sum_{\sigma \in \{\uparrow, \downarrow\}} \left( t_{\textnormal{L}}\op{c}_{-2,\sigma}^\dagger \op{c}_{-1,\sigma} + H.c. \right)\\
                                            & \quad + \sum_{\sigma \in \{\uparrow, \downarrow\}} \left( t_{\textnormal{R}}\op{c}_{1,\sigma}^\dagger \op{c}_{2,\sigma} + H.c. \right), \nonumber\\
            \op{H}_\textnormal{Leads}       &= \sum_{k=-\infty}^{-2} \left(\Delta_{\textnormal{L}} \op{c}_{k,\uparrow}^\dagger \op{c}_{k,\downarrow}^\dagger + H.c. \right) \\
                                            & \quad + \sum_{k=-\infty}^{-2}\sum_{\sigma \in \{\uparrow, \downarrow\}} \left( t_{\textnormal{L}}\op{c}_{k-1,\sigma}^\dagger \op{c}_{-k,\sigma} + H.c. \right) \nonumber \\
                                            & \quad + \sum_{k=2}^{\infty} \left(\Delta_{\textnormal{R}} \op{c}_{k,\uparrow}^\dagger \op{c}_{k,\downarrow}^\dagger + H.c. \right) \nonumber \\
                                            & \quad + \sum_{k=2}^{\infty}\sum_{\sigma \in \{\uparrow, \downarrow\}} \left( t_{\textnormal{R}}\op{c}_{k,\sigma}^\dagger \op{c}_{k+1,\sigma} + H.c. \right). \nonumber
        \end{align}
        All parameters of the model Hamiltonian are chosen real valued and positive.
        We use symmetric leads throughout this article, i.e. 
        $\Delta_{\textnormal{L}} = \Delta_{\textnormal{R}} = \Delta$ 
        and $t_{\textnormal{L}} = t_{\textnormal{R}} = t$.
        Furthermore, we will work in the wide-band limit which
        corresponds to $t_\alpha \gg t_{\textnormal{C}}$.
        Hence the results do not depend on $t_\alpha$ and $t_{\textnormal{C}}$
        independently, but only on the coupling 
        $\Gamma_\alpha = \frac{2t_{\textnormal{C}}^2}{t_\alpha}$.
        We point out that one has to solve the
        time-dependent Bogoliubov de-Gennes equation,
        which is a Schr\"odinger-like equation in the 
        electron-hole space. The corresponding
        matrices in electron-hole basis
        are written in bold-face letters. 
        
        It is useful to follow the convention introduced by Yoichiro Nambu \cite{Nambu1960} 
        and group the operators in two dimensional vectors:
        \begin{align}
            \hat{\psi}_k^\dagger &= \left( 
                                        \begin{matrix}
                                            \op{c}_{k,\up}^\dagger & \op{c}_{k,\down}
                                        \end{matrix}
                                    \right), \qquad
            \hat{\psi}_k          = \left( 
                                        \begin{matrix}
                                            \op{c}_{k,\up} \\ \op{c}_{k,\down}^\dagger
                                        \end{matrix}
                                    \right). 
        \end{align}
        The upper component represents spin up electrons,
        the lower component can be interpreted as spin down holes.
        
        For later use, it is convenient to define projections of $\Hamilton$,
        which is the matrix representation of the Hamiltonian $\op{H}$
        in Nambu space, onto
        the different subspaces. In terms of these projections, 
        the full Hamiltonian $\Hamilton(t)$ 
        can be partitioned as
        \begin{align}
            \Hamilton(t) &=             
                \left( 
                    \begin{matrix}
                        \Hamilton_{\textnormal{LL}} & \Hamilton_{\textnormal{LC}}     & 0 \\
                        \Hamilton_{\textnormal{CL}} & \Hamilton_{\textnormal{CC}}(t)  & \Hamilton_{\textnormal{CR}} \\
                        0                           & \Hamilton_{\textnormal{RC}}     & \Hamilton_{\textnormal{RR}}
                    \end{matrix}
            \right).
        \end{align}
        On the other hand, the matrix $\Hamilton_{\textnormal{CC}}(t)$ can
        be partitioned as 
        \begin{align}
            \Hamilton_{\textnormal{CC}}(t) &=             
                \left( 
                    \begin{matrix}
                        \Hamilton_{-1,-1}                  & \Hamilton_{-1,\textnormal{QD}}(t)  & 0 \\
                        \Hamilton_{\textnormal{QD},-1}(t)  & \Hamilton_{\textnormal{QD}}        & \Hamilton_{\textnormal{QD},1}(t) \\
                        0                                  & \Hamilton_{ 1,\textnormal{QD}}(t)  & \Hamilton_{1,1}
                    \end{matrix}
            \right).
            \label{eqn:Hamiltonian-HCC}
        \end{align}

    \subsection{Definition of the nonequilibrium Green's function}
\begin{figure}[htb]
            \begin{center}
                \includegraphics{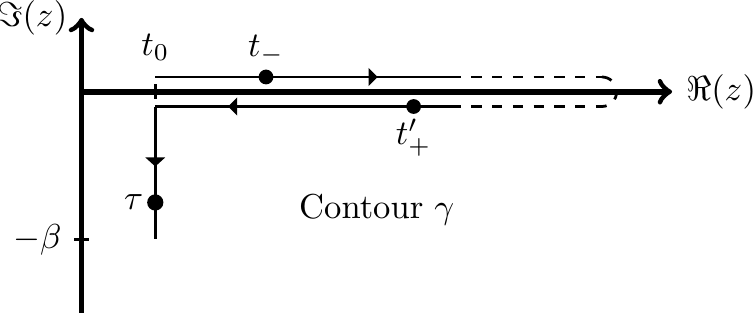}
                \caption{Sketch of the Keldysh contour $\gamma$ in the complex time plane.
                        Variables $t_{\pm}$ denote times on the upper ($-$) or lower ($+$) branch 
                        of the horizontal part. The variable $\tau$ is used for times on the 
                        vertical part. }
                \label{fig:Sketch-Keldysh-contour}
            \end{center}
        \end{figure}

        In the following sections
        we will make use of the Keldysh nonequilibrium Green's functions (NEGF).
        We shall define
        all necessary objects. For details, we refer the reader
        to the book by Stefanucci and van Leeuwen \cite{Stefanucci2013} for 
        an excellent comprehensive introduction to
        nonequilibrium Green's functions.
        The usual Keldysh contour $\gamma$ 
        is sketched in figure \ref{fig:Sketch-Keldysh-contour}.
        
        The nonequilibrium Green's function is defined as
        \begin{align}
            \G(z,z') &= -i \langle\ContourOrdering \hat{\psi}_{\textnormal{H}}(z) \hat{\psi}_{\textnormal{H}}^\dagger(z') \rangle\\
                     &= \Theta(z,z')\G^>(z,z') + \Theta(z',z)\G^<(z,z')
        \end{align}
        with the time-contour ordering operator 
        $\ContourOrdering$ and 
        the field operators $\hat{\psi}_{\textnormal{H}}(z)$ 
        and $\hat{\psi}_{\textnormal{H}}^\dagger(z)$ in the Heisenberg picture.
        The lesser and greater Green's functions for real time arguments
        are given by
        \begin{align}
            \G^<(t,t') &= i\langle\hat{\psi}^\dagger_{\textnormal{H}}(t')\hat{\psi}_{\textnormal{H}}(t) \rangle,\\
            \G^>(t,t') &= -i\langle\hat{\psi}_{\textnormal{H}}(t) \hat{\psi}_{\textnormal{H}}^\dagger(t') \rangle.
        \end{align}
        The nonequilibrium Green's functions can
        be partitioned in the same way as the
        Hamiltonian:
        \begin{align}
                \G^{\gtrless}(z,z') &=             
                \left( 
                    \begin{matrix}
                        \G^{\gtrless}_{\textnormal{LL}}(z,z')  & \G^{\gtrless}_{\textnormal{LC}}(z,z') & \G^{\gtrless}_{\textnormal{LR}}(z,z') \\
                        \G^{\gtrless}_{\textnormal{CL}}(z,z')  & \G^{\gtrless}_{\textnormal{CC}}(z,z') & \G^{\gtrless}_{\textnormal{CR}}(z,z') \\
                        \G^{\gtrless}_{\textnormal{RL}}(z,z')  & \G^{\gtrless}_{\textnormal{RC}}(z,z') & \G^{\gtrless}_{\textnormal{RR}}(z,z')
                    \end{matrix}
            \right).
            \label{eqn:partition-GF}
        \end{align}
        The same can be done for the spectral function $\SpectralFct(t,t')$.
        Similar to the partitioning of
        $\Hamilton_{\textnormal{CC}}(t)$ in equation (\ref{eqn:Hamiltonian-HCC}),
        we will use $\G^{\gtrless}_{\textnormal{QD}}(z,z')$ and $\SpectralFct_{\textnormal{QD}}(t,t')$
        in the course of this work to denote the 
        central entry of $\G^{\gtrless}_{\textnormal{CC}}(z,z')$
        and $\SpectralFct_{\textnormal{CC}}(t,t')$.

\section{Calculating the spectral function}
    \label{sec:calc-spectral-function}
    \subsection{NEGF based method}
    
    We want to calculate the large time behaviour
    of the spectral function $\SpectralFct_{\textnormal{CC}}(\omega, T)$
    with respect to $T$. We point out that
    the limit $\lim_{T\to\infty}\SpectralFct_{\textnormal{CC}}(\omega, T)$
    in general does not exist, i.e.
    $\SpectralFct_{\textnormal{CC}}(\omega, T)$
    is a nontrivial function of $T$ even for large $T$. 
    We are specifically interested
    in this $T$-dependence 
    of $\SpectralFct_{\textnormal{CC}}(\omega, T)$
    for large times, i.e. we dot not study 
    transient effects.
    We start from the definition of the spectral
    function $\SpectralFct(t,t')$ given in section \ref{sec:Introduction}:
    \begin{align}
         \SpectralFct(t,t') &= i \left[\G^> (t,t')  - \G^< (t,t')\right].
    \end{align}
    Next we use reformulate the
    Green's functions $\G^\gtrless_{\textnormal{CC}}(t,t')$
    in a way similar to Refs \onlinecite{Stefanucci2007, Stefanucci2008},
    by using the embedding self-energy 
    \begin{align}
    \SelfEnergy_\alpha(z,z') &= \Hamilton_{\textnormal{C}\alpha}\mathbf{g}_{\alpha\alpha}(z,z')\Hamilton_{\alpha\textnormal{C}},
    \end{align}
    where $\mathbf{g}_{\alpha\alpha}(z,z')$ is the nonequilibrium
    Green's function of the isolated lead $\alpha$.
    We finally arrive at:
    \begin{align}
        &\G^\gtrless_{\textnormal{CC}}(t,t') \\
              &= \sum_{\alpha, \alpha' \in \{\textnormal{L}, \textnormal{C},\textnormal{R}\}} \GRetarded_{\textnormal{C}\alpha}(t,t_0)\G^\gtrless_{\alpha \alpha'}(t_0,t_0)\GAdvanced_{\alpha'\textnormal{C}}(t_0,t') \nonumber \\
              \label{eqn:Reformulation-G-Lesser-Greater}
              & =\GRetarded_{\textnormal{CC}}(t,t_0)\G^\gtrless_{\textnormal{CC}}(t_0,t_0)\GAdvanced_{\textnormal{CC}}(t_0,t')\\
              & \qquad -i \sum_{\alpha  \in \{\textnormal{L},\textnormal{R}\}}[\GRetarded_{\textnormal{CC}}\cdot \SelfEnergy_{\alpha }^\rceil\star \G_{\textnormal{CC}}^\lceil](t,t_0)\GAdvanced_{\textnormal{CC}}(t_0,t') \nonumber\\
              & \qquad + \sum_{\alpha  \in \{\textnormal{L},\textnormal{R}\}} \GRetarded_{\textnormal{CC}}(t,t_0) [\G_{\textnormal{CC}}^\rceil \star \SelfEnergy_\alpha ^\lceil \cdot \GAdvanced_{\textnormal{CC}}](t_0,t')\nonumber \\
              & \qquad + \sum_{\alpha \in \{\textnormal{L},\textnormal{R}\}}[\GRetarded_{\textnormal{CC}}\cdot \SelfEnergy^\gtrless_\alpha  \cdot \GAdvanced_{\textnormal{CC}}](t,t') \nonumber\\
              & \qquad +\sum_{\alpha ,\alpha' \in \{\textnormal{L}, \textnormal{R}\}} [\GRetarded_{\textnormal{CC}}\cdot \SelfEnergy_{\alpha }^\rceil \star \G_{\textnormal{CC}}^\textnormal{M}\star \SelfEnergy_{\alpha'}^\lceil \cdot \GAdvanced_{\textnormal{CC}}](t,t'). \nonumber
    \end{align}
    The superscripts $\lceil$ ($\rceil$, $\textnormal{M}$) indicate that the 
    first (second, both) time argument lies on the
    vertical part of the Keldysh contour. We further 
    have employed the notation
    \begin{align}
        [\boldsymbol{A}\cdot \boldsymbol{B}](t,t') &= \int_{t_0}^\infty\,d\overline{t}\boldsymbol{A}(t,\overline{t}) \boldsymbol{B}(\overline{t}, t'),\\
        [\boldsymbol{A}\star \boldsymbol{B}](t,t') &= \int_{t_0}^{t_0-i\beta}\,d\tau \boldsymbol{A}(t,\tau) \boldsymbol{B}(\tau, t').
    \end{align}
    Multiple products are defined analogously, i.e. 
    $[\boldsymbol{A}\cdot \boldsymbol{B} \cdot \boldsymbol{C}](t,t') = [\boldsymbol{A}\cdot [\boldsymbol{B}\cdot \boldsymbol{C}]](t,t')$.  
    To investigate the 
    large-time behaviour of $\G^\gtrless_{\textnormal{CC}}(t,t')$
    with both time arguments being large, we set $t,t' > b$
    and take the limit $b\to \infty$.
    We choose the initial system at $t_0$
    such that it does not have any bound states
    and so that we can assume
    \begin{align}
        \lim_{t\to\infty}\G_{\textnormal{CC}}(t_\pm,z')    &= \lim_{t'\to\infty}\G_{\textnormal{CC}}(z,t'_\pm)=0, \\
        \lim_{t\to\infty}\SelfEnergy_\alpha(t_\pm,z')  &= \lim_{t'\to\infty}\SelfEnergy_\alpha(z,t'_\pm)=0 
    \end{align}
    for fixed $z,z'$ on the Keldysh contour. We can now
    drop all terms in equation (\ref{eqn:Reformulation-G-Lesser-Greater})
    that tend to zero as $b \to \infty$ 
    and finally arrive at:
    \begin{align}
        \G^\gtrless_{\textnormal{CC}}(t,t') &\stackrel{t,t'\to\infty}{\sim} \left[\GRetarded_{\textnormal{CC}}\cdot \SelfEnergy^\gtrless \cdot \GAdvanced_{\textnormal{CC}}\right](t,t').
        \label{eqn:Large-Time-Expression-GF}
    \end{align}
    We use the symbol $\sim$ to denote the norm convergence,
    i.e. $X(t,t') \stackrel{t,t'\to\infty}{\sim} Y(t,t')$
    means $\lim_{t,t'\to \infty}\|X(t,t')-Y(t,t')\|=0$.
    The next step consists in specifying the Hamiltonian. 
    According to the definition of the central Hamiltonian
    in equation (\ref{eqn:Hamiltonian-HCC}), we can write 
    it as
    \begin{equation}
        \Hamilton_{\textnormal{CC}}(t) = \Hamilton_{\textnormal{CC}}^0 + \mathbf{U}_+e^{i\omega_0 t} +  \mathbf{U}_-e^{-i\omega_0 t}
    \end{equation}
    with $\omega_0 = \frac{U}{2}$.
    We now assume that our system with the 
    applied bias does not have any bound states
    and that the $T$-dependence of all observables is periodic
    with frequency $\omega_0$.
    As opposed to the work of Ref \onlinecite{Stefanucci2008},
    our system has a minor history 
    and initial state dependence,
    which only shows up in a phase shift of the periodic
    observables. As we will later set the 
    initial time $t_0$ to $-\infty$,
    this phase is by design not accessible.
    This allows us to 
    represent the advanced and retarded Green's function 
    of the central region as \cite{Stefanucci2008}
    \begin{align}
        \G^{\textnormal{R}, \textnormal{A}}_{\textnormal{CC}}(t,t') &= \sum_{m\in \mathds{Z}} \int_{-\infty}^\infty \frac{\,d\omega}{2\pi}\widetilde{\G}_m^{\textnormal{R}, \textnormal{A}}(\omega)e^{-i\omega(t-t') + im\omega_0\frac{t+t'}{2}} \nonumber\\
        &= \sum_{m\in \mathds{Z}} \int_{-\infty}^\infty \frac{\,d\omega}{2\pi}\G_m^{\textnormal{R}, \textnormal{A}}(\omega)e^{-i\omega(t-t') + im\omega_0t'}.
        \label{eqn:Expansion-GF}
    \end{align}
    The equivalence of the two expansions
    can be easily checked by a variable substitution
    in the integral. We use the latter representation for convenience reasons.
    
    The embedding self-energies $\SelfEnergy_\alpha^{\textnormal{R},\textnormal{A}, \gtrless}(t,t')$ 
    depend only on the time difference $t-t'$,
    thus we can write $\SelfEnergy_\alpha^{\textnormal{R},\textnormal{A}, \gtrless}(t-t')$ 
    in terms of its Fourier transform:
    \begin{align}
        \SelfEnergy_\alpha^{\textnormal{R},\textnormal{A}, \gtrless}(t-t')   &= \int_{-\infty}^\infty \frac{\,d\omega}{2\pi} e^{-i\omega(t-t')}\SelfEnergy_\alpha^{\textnormal{R},\textnormal{A}, \gtrless}(\omega).
        \label{eqn:Expansion-Self-Energy}
    \end{align}
    At this point, it is convenient to
    set the initial time $t_0$ of the
    Keldysh contour to $-\infty$.
    We insert the expansions (\ref{eqn:Expansion-GF}) and (\ref{eqn:Expansion-Self-Energy}) into 
    equation (\ref{eqn:Large-Time-Expression-GF})
    and carry out the
    integrals,
    leading to the final expression
    \begin{align}    
    \label{eqn:Final-expression-SpecFct}
    \SpectralFct_{\textnormal{CC}}(\omega, T) \stackrel{T\to\infty}{\sim} & \frac{1}{2\pi}\sum_{\boldsymbol{m} \in \mathds{Z}^2}e^{i(m_1+m_2)\omega_0T } \\
        & \quad \cdot \G_{m_1}\left(\omega + \frac{\omega_0}{2}(m_1-m_2)\right) \nonumber \\
        & \quad \cdot \mathbf{\Gamma}\left(\omega + \frac{\omega_0}{2}(m_1-m_2)\right) \nonumber \\
        & \quad \cdot \G_{-m_2}^\dagger\left(\omega + \frac{\omega_0}{2}(m_1-m_2)\right) \nonumber ,        
    \end{align}
    \begin{align}    
    \SpectralFct_{\textnormal{CC,DC}}(\omega) &= \lim_{\tilde{t}\to\infty} \frac{\omega_0}{2\pi} \int_{\tilde{t}}^{\tilde{t}+\frac{2\pi}{\omega_0}}\,dT \SpectralFct_{\textnormal{CC}}(\omega, T) \\
    & = \frac{1}{2\pi}\sum_{m \in \mathds{Z}} \G_{m}\left(\omega + \omega_0m\right) \\
    & \qquad \mathbf{\Gamma}\left(\omega + \omega_0m \right) \G_{m}^\dagger\left(\omega + \omega_0m \right) \nonumber ,        
    \end{align}
    with 
    \begin{align}
        \mathbf{\Gamma}_\alpha(\omega)  &= i \left[ \SelfEnergy_\alpha^{\textnormal{R}}(\omega) - \SelfEnergy_\alpha^{\textnormal{A}}(\omega)\right],\\
        \mathbf{\Gamma}(\omega)         &= \mathbf{\Gamma}_\textnormal{L}(\omega)+\mathbf{\Gamma}_\textnormal{R}(\omega),\\
        \G_m(\omega)                    &= \GRetarded_m(\omega-m\omega_0) = \left[\GAdvanced_{-m}(\omega)\right]^\dagger.
    \end{align}
    The matrices $\G_m(\omega)$ are
    calculated by using a scheme
    proposed by Stefanucci \textit{et al.}. \cite{Stefanucci2008}

    The analytic expression for the embedding self-energies 
    $\SelfEnergy^{\textnormal{R}/\textnormal{A}}_\alpha(\omega)$ 
    of our model Hamiltonian 
    reads \cite{Perfetto2009}:
    \begin{align}
     \SelfEnergy^{\textnormal{R}/\textnormal{A}}_\alpha(\omega) 
        &= \lim_{\eta \searrow 0}
            \left(
                \begin{matrix}
                    \tilde{m}_\alpha(\omega \pm i \eta) & \tilde{d}_\alpha(\omega \pm i \eta)\\
                    \tilde{d}_\alpha(\omega \pm i \eta) & \tilde{m}_\alpha(\omega \pm i \eta)
                \end{matrix}
            \right),\\
        \tilde{m}_\alpha(z) &= \frac{z}{2}\frac{\sqrt{\Delta_\alpha^2-z^2} - \sqrt{\Delta_\alpha^2-z^2+4t_\alpha^2}}{\sqrt{\Delta_\alpha^2-z^2}},\\
        \tilde{d}_\alpha(z) &= \frac{\Delta_\alpha}{2}\frac{\sqrt{z^2-\Delta_\alpha^2-4t_\alpha^2} - \sqrt{z^2-\Delta_\alpha^2}}{\sqrt{z^2-\Delta_\alpha^2}}.
    \end{align}
    In the numerical implementation, one has
    to use a finite value of $\eta$ which
    will be stated in the results
    for the sake of completeness.
    
    The remaining
    problem is the numerical evaluation of 
    the infinite sum in equation (\ref{eqn:Final-expression-SpecFct}).
    We do this by truncation
    at some large value $m_\textnormal{max}$ 
    i.e. the sum runs over
    $\|\boldsymbol{m}\|<m_\textnormal{max}, \boldsymbol{m} \in \mathds{Z}^2$.
    Of course, the convergence of the results with 
    respect to $m_\textnormal{max}$ 
    has to be checked carefully.
    
    \subsection{Time propagation method}
    An alternative way to calculate the nonequilibrium Green's  
    functions $\G^{\gtrless}(t,t')$ is to use 
    propagated single particle wave functions
    $\psi_{q}(t) = (u_{q}(t), v_{q}(t))^T$. 
    
    We start in the ground state
    at $t=0$ and carry out 
    a time propagation only in the central region, i.e. we solve
    \begin{align}
        \left[i\partial_t - \Hamilton_{\textnormal{CC}}(t) \right]&\psi_{q,\textnormal{C}}(t) = \\
         & \sum_{\alpha \in \{\textnormal{L},\textnormal{R}\}}  \int_0^t\,dt' \SelfEnergy^{\textnormal{R}}_\alpha(t,t') \psi_{q,\textnormal{C}}(t')  \nonumber \\
         &\quad + \sum_{\alpha \in \{\textnormal{L},\textnormal{R}\}}  \Hamilton_{\textnormal{C}\alpha}\mathbf{g}_{\alpha\alpha}^\textnormal{R}(t,0)\psi_{q,\alpha}(0) . \nonumber
    \end{align}
    We refer the reader to the 
    work of
    Stefanucci \textit{et al.} \cite{Stefanucci2010}
    for details of the propagation scheme.
 
    Having the single particle wave functions $\psi_q(t)$ at hand,
    we can calculate the 
    spectral function $ \SpectralFct(t,t')$ as
    \begin{align}
        \left[\SpectralFct(t,t')\right]_{kl} &=
            \sumint_q  
            \left(
                \begin{matrix}
                    u_{q,k}(t)\left[u_{q,l}(t')\right]^\star & u_{q,k}(t)\left[v_{q,l}(t')\right]^\star\\
                    v_{q,k}(t)\left[u_{q,l}(t')\right]^\star & v_{q,k}(t)\left[v_{q,l}(t')\right]^\star
                \end{matrix}
            \right),
        \label{eqn:Def-SpecFct-Wavefunctions}
    \end{align}
    where $\sumint_q$ stands for the
    integration over scattering states
    and the summation over bound 
    states. 
    The spectral function $\SpectralFct(\omega, T)$
    can then be calculated using equation (\ref{eqn:Def-SpecFct-Energy-Time})
    and a numerical Fourier transform on
    an equidistant grid in the time domain.
    This formulation has the advantage that
    one can obtain $\SpectralFct(\omega, T)$
    for all times $T$ and not just the large 
    time behaviour. 
    But, the NEGF based method is 
    computationally faster and 
    has a better numerical accuracy.
    Hence, it is used whenever applicable.
    Of course, both methods yield
    the same results
    in the large-time limit.
    
        \begin{figure*}[htb]
            \begin{center}
                \includegraphics{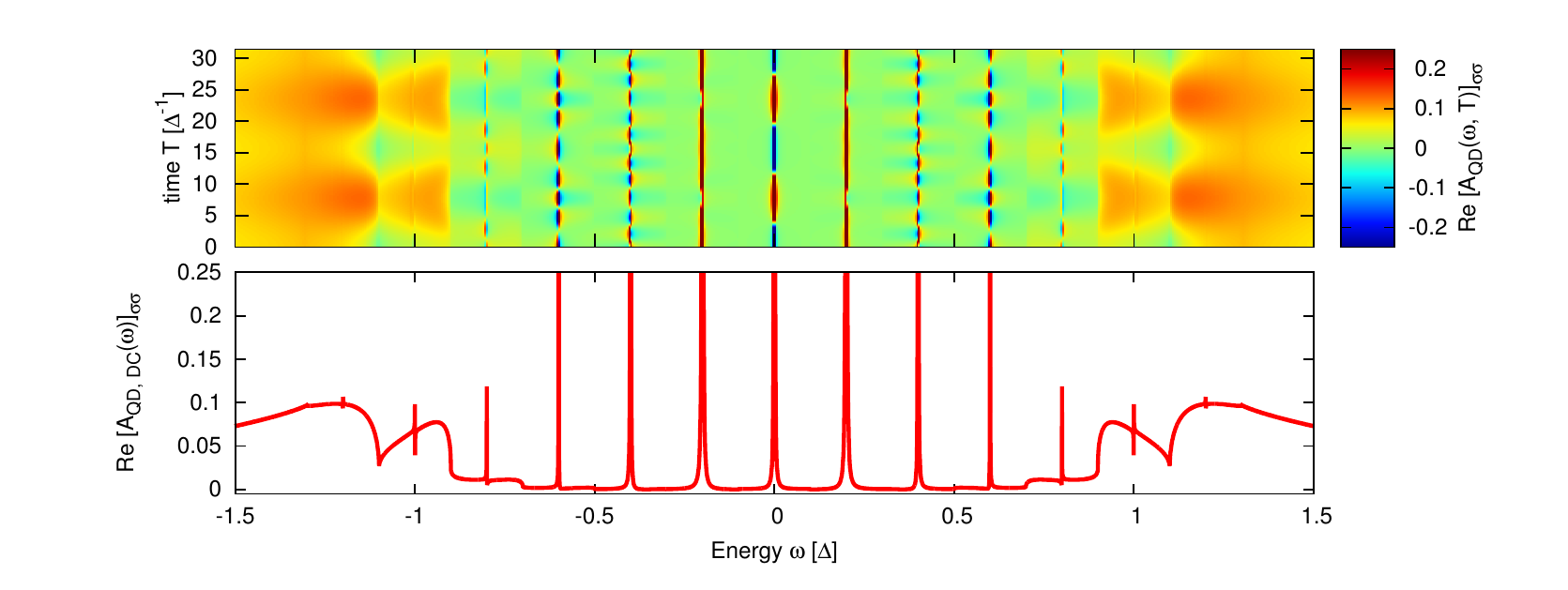}
                \caption{Top: The spectral function 
                         $[\SpectralFct_{\textnormal{QD}}(\omega, T)]_{\sigma\sigma}$.
                         The results for $\sigma=\ \uparrow$ and 
                         $\sigma=\ \downarrow$ are identical. 
                         Bottom: The time-averaged spectral function 
                         $[\SpectralFct_{\textnormal{QD},\textnormal{DC}}(\omega)]_{\sigma\sigma}$.
                         The bias has been switched 
                         on in the past. The imaginary part
                         of the spectral function vanishes. 
                         The peaks have a finite width and height.
                         The parameters are: $\Gamma_\alpha=0.5\Delta$ and
                         $U_{\textnormal{L}} = -U_{\textnormal{R}} = 0.1\Delta, \eta = 10^{-9}\Delta, m_\textnormal{max}=20$.}
                \label{fig:Result-Pure}
            \end{center}
        \end{figure*}  
        
\section{Results - Occurrence of negative values}
    \label{sec:results-negative-values}
    \subsection{Spectral Function for a constant bias}
   
        We start with the presentation
        of an example
        demonstrating that a simple
        probability interpretation
        of the spectral function is not possible.
        Figure \ref{fig:Result-Pure} shows the calculated
        diagonal elements of the
        spectral function $\SpectralFct_{\textnormal{QD}}(\omega, T)$
        as well as its time average.
        Both are computed with the help of the NEGF based method.
        The bias has been applied in the past. 
        The whole structure is periodic in time 
        according to the Josephson effect with 
        frequency $\omega_{\textnormal{J}} = \frac{2eU}{\hbar}$.
        Both edges of the superconducting gap structure 
        are split up. The new ones are located 
        at $\Delta\pm\frac{U}{2}$ and $-\Delta\pm\frac{U}{2}$.
        
        Inside the gap,
        there is a peak structure with a spacing between the peaks of the bias $U$. 
        In the plot of the time resolved 
        spectral function $\SpectralFct_{\textnormal{QD}}(\omega, T)$,
        we further observe small areas with negative values, mostly inside the gap.
        These findings contradict the interpretation as 
        a time-dependent density of states, 
        which has to be non-negative.
        The effect is absent in
        the time-averaged spectral function
        $\SpectralFct_{\textnormal{QD}, \textnormal{DC}}(\omega)$:
        It is non-negative for all energies $\omega$.
        This phenomenon has already been
        found in a study on
        metallic rings \cite{Arrachea2002}, but was not 
        further investigated.
        To the best of our knowledge,
        negative values of $\SpectralFct(\omega, T)$
        did not occur in
        all other studies of
        this quantity.

        We investigate this issue 
        and provide a reasonable
        explanation incorporating these 
        negative areas. The diagonal of the spectral 
        function $\SpectralFct(\omega, T)$
        is an analogue of the Wigner function. 
        The latter is defined as \cite{Wigner1932}
        \begin{align}
            P(x,p) &= \frac{1}{2\pi\hbar}\int_{-\infty}^\infty\,dy e^{-ipy/\hbar} \psi\left(x+\frac{y}{2}\right)\left[\psi\left(x-\frac{y}{2}\right)\right]^\star\\
                   &= \frac{1}{2\pi\hbar}\int_{-\infty}^\infty\,dq e^{iqx/\hbar} \psi\left(p+\frac{q}{2}\right)\left[\psi\left(p-\frac{q}{2}\right)\right]^\star.
        \end{align}
        It links the quantum mechanical wave functions $\psi(x)$ and $\psi(p)$ to 
        a phase space distribution $P(x,p)$. 
        But, $P(x,p)$ can have small negative areas \cite{Schleich2001}.
        Hence, the simple probability interpretation
        of the Wigner function $P(x,p)$ 
        is not possible. 
        In fact, it is not unusual that $P(x,p)$ has negative areas.
        This issue has been solved \cite{Stenholm1980} by taking into account the
        uncertainty principle:
        \begin{align}
            \sigma_{x}\sigma_{p} &\geq \frac{\hbar}{2}, \\
            \sigma_{x}&=\sqrt{\langle \hat{x}^{2} \rangle-\langle \hat{x}\rangle ^{2}},\\
            \sigma_{p}&=\sqrt{\langle \hat{p}^{2} \rangle-\langle \hat{p}\rangle ^{2}}.
        \end{align}
    
        \begin{figure*}[htb]
            \begin{center}
                \includegraphics{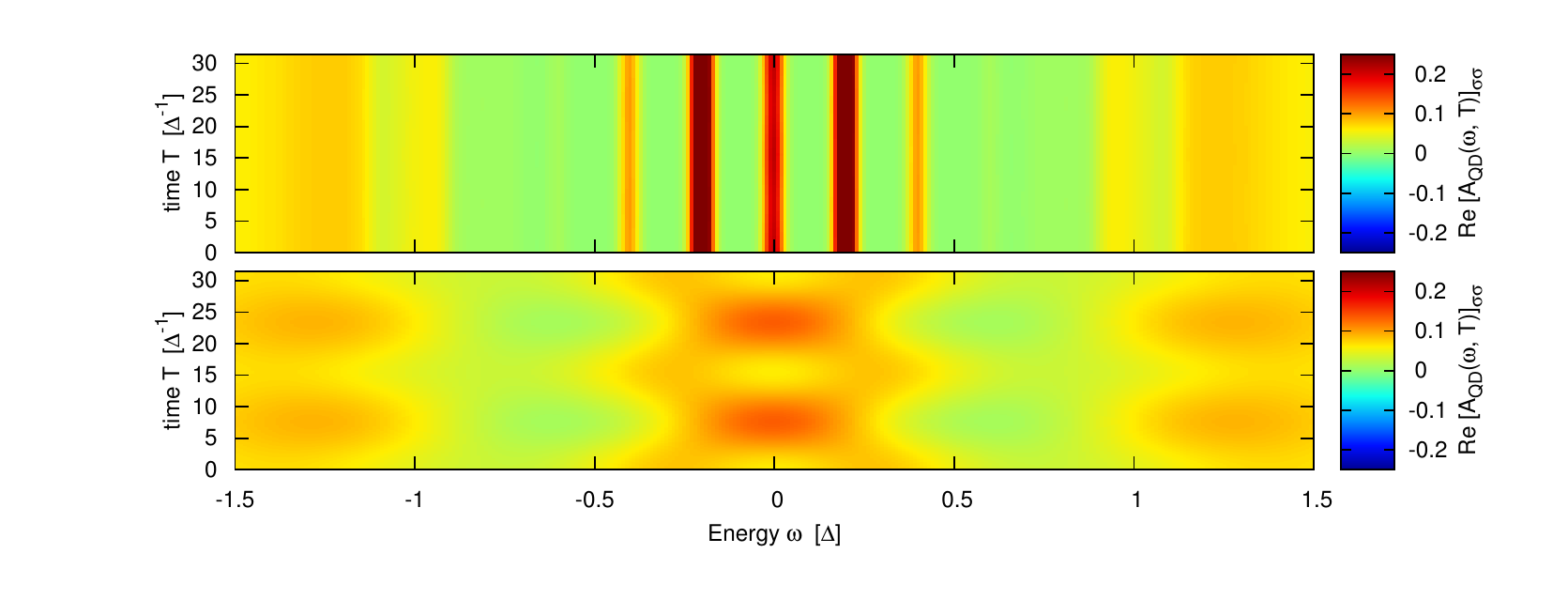}
                \caption{A plot of the convoluted spectral function
                        $\left[\widetilde{\SpectralFct}_{\textnormal{QD}}(\omega, T)\right]_{\sigma\sigma}$. It can be compared 
                        to the original one 
                        $\left[\SpectralFct_{\textnormal{QD}}(\omega, T)\right]_{\sigma\sigma}$
                        in figure \ref{fig:Result-Pure} (top).
                        As expected, there are no more negative areas.
                        The standard deviations are: 
                        $\sigma_\omega = \frac{1}{50}\Delta$, $\sigma_T=25\Delta^{-1}$ (top)
                        and $\sigma_\omega = \frac{1}{5}\Delta$, $\sigma_T=\frac{5}{2}\Delta^{-1}$ (bottom).
                        Both examples fulfill $\sigma_\omega \sigma_T = \frac{1}{2}$.
                        All other parameters are identical to those in figure     
                        \ref{fig:Result-Pure}.}
                \label{fig:Result-Convoluted}
            \end{center}
        \end{figure*}
        
        This means that one cannot determine 
        momentum and position with arbitrary precision
        at the same time
        and hence, a phase space distribution
        $P(x,p)$ with arbitrary sharp values of $x$ and 
        $p$ does not make sense in the quantum world.
        In order to achieve a measurable quantity one should
        take a weighted average 
        of $P(x,p)$ in some region,
        i.e. one should 
        for instance investigate
        \begin{align}
            & \widetilde{P}(x,p)             = \iint_{-\infty}^{\infty} \,dx'\,dp' P(x',p') M_{\sigma_x, \sigma_p}(x-x', p-p'),\\
            & M_{\sigma_x, \sigma_p}(x, p)   = \frac{\hbar}{2\pi\sigma_{x}\sigma_{p}} e^{- \frac{x^2}{2\sigma_x^2} - \frac{p^2}{2\sigma_p^2} }
        \end{align}
        rather than $P(x,p)$. 
        It has been proven that $\widetilde{P}(x,p)$
        is non-negative provided that $\sigma_{x}, \sigma_{p}>0$ 
        and $\sigma_{x}\sigma_{p}\ge \frac{\hbar}{2}$ \cite{Cartwright1975}.
        This allows a
        probability interpretation of 
        $\widetilde{P}(x,p)$ including 
        the momentum and position uncertainties.

    \subsection{Recovering the positive semidefiniteness - The convoluted spectral function }

        The formulation of the spectral function 
        $\SpectralFct(\omega, T)$ stated in equation (\ref{eqn:Def-SpecFct-Wavefunctions})
        is very similar 
        to the Wigner function $P(x,p)$.
        The time $T$ is the analogue of the momentum $p$
        and the frequency $\omega$ corresponds to the position $x$.
        Hence it is not surprising to see
        that the spectral function 
        $\SpectralFct(\omega, T)$ can have small
        negative areas. The time-energy uncertainty relation
        plays the same role for $\SpectralFct(\omega, T)$
        as does the $x$-$p$ uncertainty relation for $P(x,p)$.
        Like for the Wigner
        function, we can verify
        the non-negativity of the 
        convoluted spectral function:
        \begin{align}
            0 & \le \left[\widetilde{\SpectralFct}_{\textnormal{QD}}(\omega, T)\right]_{\sigma \sigma} \\ 
            & = \iint_{-\infty}^{\infty} \,d\omega'\,dT' \left[\SpectralFct_{\textnormal{QD}}(\omega', T')\right]_{\sigma \sigma} \\
            & \qquad \qquad \qquad  \cdot M_{\sigma_\omega, \sigma_T}(\omega-\omega', T-T') \nonumber ,
        \end{align}
        with
        \begin{align}
            M_{\sigma_\omega, \sigma_T}(\omega, T)   &= \frac{1}{2\pi\sigma_{\omega}\sigma_{T}} e^{- \frac{\omega^2}{2\sigma_\omega^2} - \frac{T^2}{2\sigma_T^2}},
        \end{align}
        provided that $\sigma_\omega, \sigma_T > 0$ and $\sigma_\omega \sigma_T \ge \frac{1}{2}$.
        The proof is carried out in the appendix \ref{app:proof}.

        We further point out that $\SpectralFct_{\textnormal{QD}}(\omega, T)$ and $\widetilde{\SpectralFct}_{\textnormal{QD}}(\omega, T)$
        are normalized such that
        \begin{align}
            1 &= \int_{-\infty}^\infty \,d\omega \left[\SpectralFct_{\textnormal{QD}}(\omega, T)\right]_{\sigma\sigma} \\
              &= \int_{-\infty}^\infty \,d\omega \left[\widetilde{\SpectralFct}_{\textnormal{QD}}(\omega, T)\right]_{\sigma \sigma}
        \end{align}
        for $\sigma \in \{\uparrow, \downarrow\}$ and all times $T$. Hence we conclude that 
        $\widetilde{\SpectralFct}_{\textnormal{QD}}(\omega, T)$
        is a probability density function
        with respect to $\omega$
        and allows a direct
        interpretation as a $T$-dependent
        density of states. We note in passing 
        that a probability interpretation
        referring to the variable $T$ is generally 
        not possible because there is no 
        normalization condition with respect to $T$.
        
        Two different convolutions of the example in figure \ref{fig:Result-Pure}
        are shown in figure \ref{fig:Result-Convoluted}. 
        There are no more negative areas as expected.         
        Hence, the spectral function $\SpectralFct(\omega, T)$
        has to be viewed as a quasiprobability density function
        and only the convoluted spectral function
        $\widetilde{\SpectralFct}_{\textnormal{QD}}(\omega, T)$
        should be used for direct comparisons with experiments.
        Furthermore, the two examples show that it is 
        impossible to resolve the peak structure 
        inside the gap and observe the periodicity
        due to the Josephson effect simultaneously.
        The spacing of the peak structure is
        $U$ and time periodicity is $\frac{\hbar}{e}\frac{\pi}{U}$.
        Hence, one can not resolve both
        features simultaneously without 
        violating the time-energy uncertainty relation.

        Having this interpretation at hand, 
        we are able to explain the structure inside 
        the gap. It is created by particles
        that cross the gap 
        with the help of Andreev reflections.
        This charge transfer mechanism has been
        used to explain the subharmonic 
        gap structure in several works\cite{Yeyati1997,Johansson1999}.
        The negative areas can be viewed as 
        interference effects and reveal the 
        quantum nature of the particles involved.
        
        \begin{figure*}[htb]
            \begin{center}
                \includegraphics[scale=1]{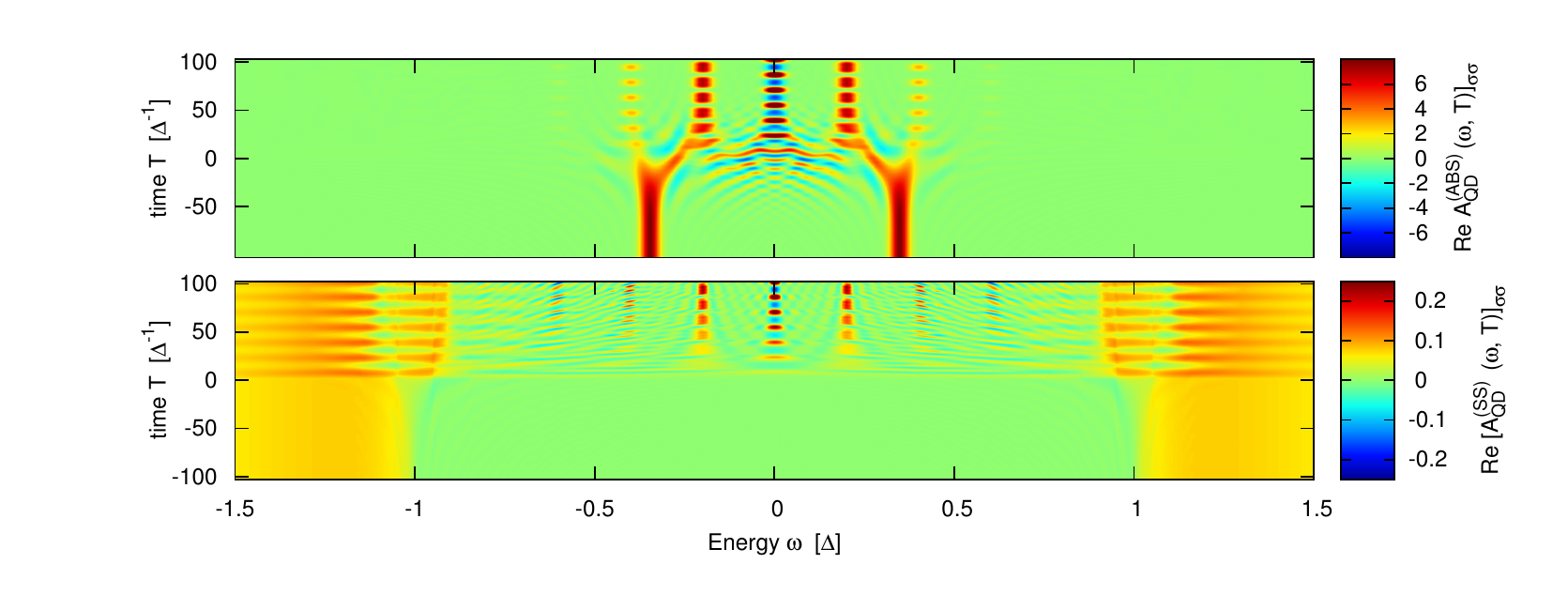}
                \caption{Decomposition of the full spectral function $\SpectralFct_{\textnormal{QD}}(\omega, T)$
                         into contributions of the Andreev bound states 
                         $\SpectralFct^{(\textnormal{ABS})}_{\textnormal{QD}}(\omega, T)$ (top, convoluted in frequency space)  
                         and the scattering states $\SpectralFct^{(\textnormal{SS})}_{\textnormal{QD}}(\omega, T)$ (bottom). The bias is switched on at $t=0$. 
                         The parameters are: $\Gamma_\alpha=0.5\Delta, U_{\textnormal{L}}(t) = -U_{\textnormal{R}}(t) = 0.1\Delta^{-1}\Theta(t)$.}
                 \label{fig:Switch-on-bias}
            \end{center}
        \end{figure*}
        
\section{Results - Visualization of switching processes}

        \label{sec:results-switching-processes}

        We can also visualize switching effects 
        with the help of the spectral function.
        We prepare the system in the ground state for $t<0$ and switch on 
        the bias at $t=0$.
        The spectral function is split up into contributions
        of Andreev bound states (labeled $\SpectralFct^{(\textnormal{ABS})}_{\textnormal{QD}}(\omega, T)$) 
        and scattering states (labeled $\SpectralFct^{(\textnormal{SS})}_{\textnormal{QD}}(\omega, T)$).
        All results of this subsection 
        are calculated using the formulation 
        of equation (\ref{eqn:Def-SpecFct-Wavefunctions})
        for the spectral function. The Fourier transform 
        with respect to $t-t'$ is done on a grid. 
        We therefore demand for 
        $\SpectralFct(t,t') \to 0$
        as $|t-t'| \to \infty$.
 
    \subsection{Handling bound states}

        In contrast to the previous results,
        we now have to take Andreev bound states
        into account. These underline the
        importance of the previously
        discussed convolutions
        in a second way.
        
        Consider a system with a single bound state (BS) whose
        energy $\epsilon_{\textnormal{BS}}(t)$ changes 
        adiabatically. Its spectral function
        $\SpectralFct^{(\textnormal{BS})}(t,t')$
        is then proportional to
        \begin{align}
           \SpectralFct^{(\textnormal{BS})}(t,t') & =  \psi^{(\textnormal{BS})}(t) \left[\psi^{(\textnormal{BS})}(t')\right]^\star\\
                                                  & \sim e^{+i \epsilon_{\textnormal{BS}}(t) t}e^{-i \epsilon_{\textnormal{BS}}(t') t'}.
        \end{align}
        Hence we can conclude $\SpectralFct^{(\textnormal{BS})}(t,t') \not\to 0$
        as $|t-t'| \to \infty$. If we then calculate the
        Fourier transform of $\SpectralFct^{(\textnormal{BS})}(t,t')$
        with respect to $\tau = t-t'$,
        the full past and future relative to the time $T=\frac{t+t'}{2}$
        influences the value of $\SpectralFct^{(\textnormal{BS})}(\omega, T)$.
        The spectral function $\SpectralFct^{(\textnormal{BS})}(\omega, T)$
        is not just a single peak at $\epsilon_{\textnormal{BS}}(T)$
        as one would expect since everything changes adiabatically.
        
        This is in contrast to the most common situation where 
        $\SpectralFct(t,t') \to 0$
        as $|t-t'| \to \infty$. In this case, only the wave functions
        at times in a neighborhood of $T=\frac{t+t'}{2}$
        have an influence on the value of $\SpectralFct(\omega, T)$.
        Besides, this allows us to calculate the
        Fourier transform numerically on a grid.
        
        The most natural way to recover 
        the desired behaviour 
        even in the presence of bound states is to 
        enforce the decay $\SpectralFct(t,t') \to 0$
        as $|t-t'| \to \infty$ by hand. 
        In this way, one mimics
        a finite lifetime of
        the bound states.
        We now show that this is automatically done
        by the convolution
        in frequency space presented above.
               
        Using the convolution theorem, 
        the convolution in frequency 
        space can be reformulated as:
        \begin{align}
            &\frac{1}{\sqrt{2\pi}\sigma_\omega}
            \int_{-\infty}^\infty \,d´\omega' \SpectralFct(\omega', T)
            e^{-\frac{(\omega-\omega')^2}{2\sigma_\omega}} \\
            &\quad = \int_{-\infty}^\infty \,d\tau e^{i\omega \tau} e^{-\frac{\sigma_\omega^2}{2}\tau^2}
              \SpectralFct\left(T+\frac{\tau}{2}, T-\frac{\tau}{2}\right). \nonumber
        \end{align}
        Hence, the convolution of the spectral 
        function adds the desired decay 
        $e^{-\frac{\sigma_\omega^2}{2}(t-t')^2}\SpectralFct(t,t') \to 0$
        as $|t-t'| \to \infty$ 
        even for bound states.
        Simultaneously, it broadens all sharp frequency peaks
        of $\SpectralFct(\omega, T)$.
     
        Since the 
        convolutions smoothen the plots,
        we only apply it
        in frequency space 
        to all calculations covering Andreev bound states.
        The time convolution is not performed 
        in any of the following plots
        in order to obtain sharper structures.
        The value of $\sigma_\omega$ is set to $\sqrt{2}\cdot 0.025\Delta$ 
        for all plots covering bound states in this section.
        
    \subsection{Switching on the bias}   

        Figure \ref{fig:Switch-on-bias} shows an example of a switching process. 
        The top part shows the contribution of 
        the Andreev bound states $\SpectralFct^{(\textnormal{ABS})}_{\textnormal{QD}}(\omega, T)$
        to the spectral function, the lower part the one
        of the scattering states $\SpectralFct^{(\textnormal{SS})}_{\textnormal{QD}}(\omega, T)$.
        The bias is turned on at $t=0$.
        
        In the lower part of figure \ref{fig:Switch-on-bias}, the structure 
        already observed in 
        figure \ref{fig:Result-Pure} (top) starts to
        develop after the bias is switched on at $t=0$. 
        Simultaneously, the Andreev bound 
        states move gradually into the leads. This 
        can hardly be seen in the top figure since 
        the decay rate in the chosen example 
        is too slow compared to the 
        plotted time range. Their
        contribution to the spectral function
        will eventually be gone.
        
        We observe effects of the bias already at times $t<0$
        which hints at a violation of causality of the spectral
        function. But, in order to compare the results with an experiment,
        one has to apply convolutions as explained above.
        This resolves the issue 
        for the examples investigated here.
        Whether this is generally true is currently unknown.
    
    \subsection{Andreev bound states under the influence of a bias pulse}
        \begin{figure}[htb]
            \begin{center}
                \includegraphics[scale=1]{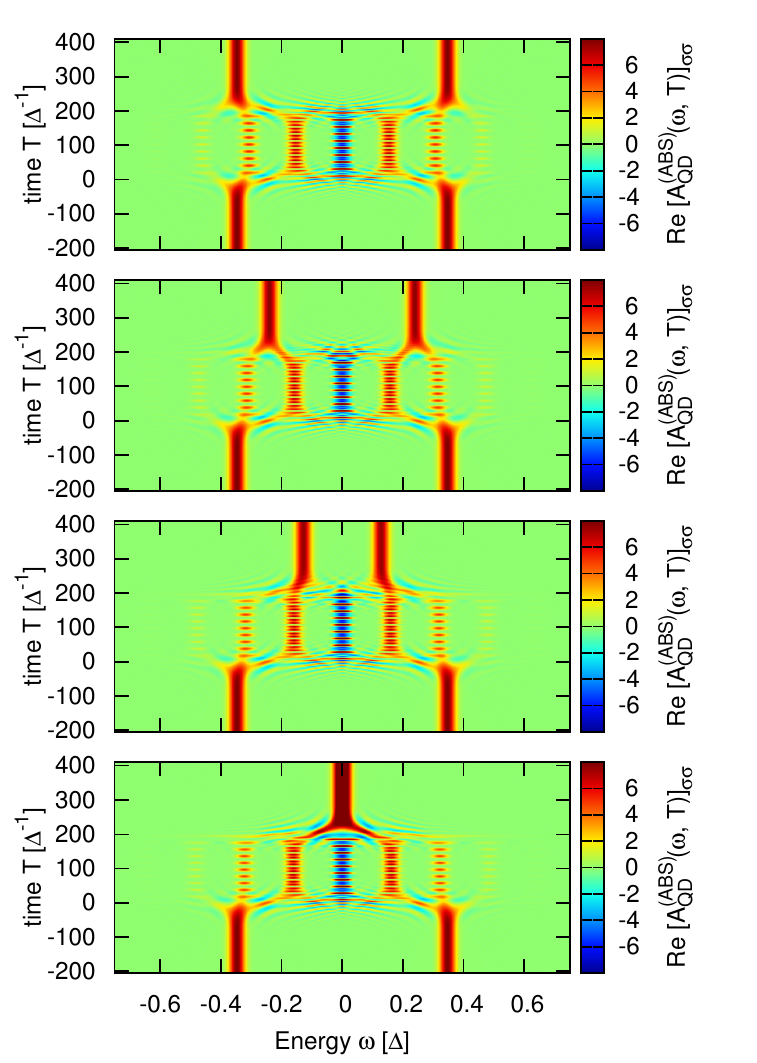}
                \caption{The spectral function of the Andreev bound states under the influence of a bias pulse from 
                         $t_0$ to $t_1$. The pulse is chosen such that the accumulated phase is
                         $\Delta \chi=\frac{2e}{\hbar}\int_{t_0}^{t_1}U(t')\,dt' = (20+x)\pi, x \in \{0,0.5,0.75,1\}$ (from top to bottom),
                         $t_0=0, t_1=204.8\Delta^{-1}$. The coupling is $\Gamma_\alpha=0.5\Delta$.
                         The plots show results convoulted in frequency space with $\sigma_\omega = \sqrt{2} \cdot 0.025\Delta$.}
                \label{fig:Switch-ABS}
            \end{center}
        \end{figure}
        
        The reformation of the Andreev bound states 
        after a bias pulse is shown in figure \ref{fig:Switch-ABS}.
        Their energy after the bias depends on the accumulated
        phase $\chi(t) = \frac{2e}{\hbar}\int_0^t\,dt'U(t')$.
        In this situation, the location of the Andreev bound states
        as a function of the phase difference $\chi$
        can be calculated by solving \cite{Stefanucci2010}
        \begin{align}
            0 &= x^2\left(1+\frac{\gamma}{\sqrt{1-x^2}}\right)-\frac{\gamma^2}{1-x^2}\frac{1+\cos \chi}{2}
        \end{align}
        with $x = \frac{\omega}{\Delta}$
        and $\gamma = \frac{\Gamma}{\Delta}$.        
        
        After the pulse, the system does not 
        evolve towards the ground state again, but
        shows non-decaying oscillations. The frequency is
        the energy difference 
        of the newly formed Andreev bound states \cite{Stefanucci2010}.

\section{Conclusions}
    \label{sec:conclusion}
    In this paper we have investigated 
    the spectral function $\SpectralFct(\omega, T)$
    for a Josephson junction in the 
    presence of a bias. 
    It turns out that 
    this quantity can occasionally have
    negative values implying that the
    interpretation as a time-dependent 
    density of states is problematic. 
    Viewing the spectral function $\SpectralFct(\omega,T)$
    as the time-energy 
    analogue of the Wigner function $P(x,p)$
    provides a way out. The 
    latter is well known to
    have negative values. 
    This issue is solved 
    by looking at averages which then are 
    strictly non-negative. 
    We do the same for 
    the spectral function, which then 
    allows a physically meaningful interpretation as
    a time-dependent 
    density of states. It
    provides useful insights into
    the internal changes of the quantum dot,
    which are illustrated by three 
    examples differing 
    in the way the bias is turned on:
    First, the bias is switched on in the past 
    and one looks at the asymptotic
    behaviour for large times.
    Second, we look at the transients
    when switching the bias in
    a step-like fashion.
    Third, the reformation
    of Andreev bound states is
    shown under the influence of 
    a bias pulse.
    All three examples
    show a non-trivial
    time-dependence which 
    can be nicely understood by 
    interpreting the 
    spectral function $\SpectralFct(\omega, T)$
    as a time-dependent density of states.

\appendix

\section{Proof}
    \label{app:proof}
    The goal of this section is to prove the 
    following relation:
    \begin{align}
        0 & \le \left[\widetilde{\SpectralFct}_{\textnormal{QD}}(\omega, T)\right]_{\sigma \sigma} \\ 
        & = \iint_{-\infty}^{\infty} \,d\omega'\,dT' \left[\SpectralFct_{\textnormal{QD}}(\omega', T')\right]_{\sigma \sigma} \\
        & \qquad \qquad \qquad  \cdot M_{\sigma_\omega, \sigma_T}(\omega-\omega', T-T') \nonumber ,
    \end{align}
    for $\sigma_\omega, \sigma_T > 0$, $\sigma_\omega \sigma_T \ge \frac{1}{2}$
    and the Gaussian kernel $M_{\sigma_\omega, \sigma_T}(\omega, T)$
    defined as
    \begin{align}
        M_{\sigma_\omega, \sigma_T}(\omega, T)   &= \frac{1}{2\pi\sigma_{\omega}\sigma_{T}} e^{- \frac{T^2}{2\sigma_T^2} - \frac{\omega^2}{2\sigma_\omega^2}}.
    \end{align}
    
    The proof follows a previous work of Cartwright \cite{Cartwright1975}.
    We will make use of the subsequent relations:
    \begin{align}
        \frac {1}{\sigma\sqrt{2\pi}} \int_{-\infty}^\infty \,dx e^{-\frac {1}{2} \left(\frac{x-\mu}{\sigma}\right)^2}                 &= 1,\\
        \frac {1}{\sigma\sqrt{2\pi}} \int_{-\infty}^\infty \,dx e^{ i\omega x} e^{-\frac {1}{2} \left(\frac{x-\mu}{\sigma}\right)^2}  &= e^{-\frac{\sigma^2\omega^2}{2} + i\mu\omega}.
    \end{align}
    We furthermore define
    \begin{align}
        f_{\sigma_\omega, \sigma_T}(x, \omega, T) = \frac{1}{\sqrt{2\pi}} e^{\frac{ i\omega}{2}x -\frac{\sigma_\omega^2x^2}{2}- \frac{x^2 - Tx}{2\sigma_T^2}}.
    \end{align}
    The proof  is carried out for 
    $\left[\widetilde{\SpectralFct}_{\textnormal{QD}}(\omega, T)\right]_{\uparrow \uparrow}$,
    its generalization for the component $\downarrow$ as well as 
    for other site positions of the tight binding chain is 
    straightforward.

    \begin{widetext}
        \begin{align}
            \left[\widetilde{\SpectralFct}_{\textnormal{QD}}(\omega, T)\right]_{\uparrow\uparrow} & = \iint_{-\infty}^{\infty} \,d\omega'\,dT' \left[\SpectralFct_{\textnormal{QD}}(\omega', T')\right]_{\sigma \sigma} 
                M_{\sigma_\omega, \sigma_T}(\omega-\omega', T-T')\\
            & = \frac{1}{4\pi^2\sigma_{\omega}\sigma_{T}} \iiint_{-\infty}^{\infty} \,d\omega'\,dT' \,d\tau' \left[\SpectralFct_{\textnormal{QD}}\left(T' + \frac{\tau'}{2},T'-\frac{\tau'}{2}\right)\right]_{\uparrow\uparrow} e^{i\omega' \tau' }  e^{- \frac{(T-T')^2}{2\sigma_T^2} - \frac{(\omega-\omega')^2}{2\sigma_\omega^2}}\\
            & = \frac{\sqrt{2\pi}}{4\pi^2\sigma_{T}} \iint_{-\infty}^{\infty} \,dT' \,d\tau' \left[\SpectralFct_{\textnormal{QD}}\left(T' + \frac{\tau'}{2},T'-\frac{\tau'}{2}\right)\right]_{\uparrow\uparrow} e^{-\frac{\sigma_\omega^2\tau'^2}{2} + i\tau'\omega}  e^{- \frac{(T-T')^2}{2\sigma_T^2}}\\
            & \underset{y = T'-\frac{\tau'}{2}}{\overset{x = T'+\frac{\tau'}{2}}{=}}
                \frac{2}{2\pi\sqrt{2\pi}\sigma_{T}} \iint_{-\infty}^{\infty} \,dx \,dy \left[\SpectralFct_{\textnormal{QD}}\left(x, y\right)\right]_{\uparrow\uparrow} e^{-\frac{\sigma_\omega^2(x-y)^2}{2} + i(x-y)\omega}  e^{- \frac{(T-\frac{x+y}{2})^2}{2\sigma_T^2}}\\
            & =
            \frac{2 e^{- \frac{T^2}{2\sigma_T^2}} }{\sqrt{2\pi}\sigma_{T}} \iint_{-\infty}^{\infty} \,dx \,dy \sumint_q 
                u_{q,\textnormal{QD}}(x) \left[u_{q,\textnormal{QD}}(x)\right]^\star 
                f(x, \omega, T) \left[f(y, \omega, T)\right]^\star e^{\sigma_\omega^2 xy - \frac{xy}{4\sigma_T^2}}\\
            & = \underbrace{\frac{2 e^{- \frac{T^2}{2\sigma_T^2}} }{\sqrt{2\pi}\sigma_{T}}}_{> 0,\textnormal{ since }\sigma_T >0} \sumint_q  \sum_{k=0}^\infty
            \underbrace{\frac{(\sigma_\omega^2-\frac{1}{4\sigma_T^2})^k}{k!}}_{\ge 0, \textnormal{ since } \sigma_\omega \sigma_T \ge \frac{1}{2}}\\
            & \nonumber \qquad \qquad 
            \underbrace{\left[\int_{-\infty}^\infty \,dx\  u_{q,\textnormal{QD}}(x) f(x, \omega, T) x^k \right] \left[\int_{-\infty}^\infty \,dy \ u_{q,\textnormal{QD}}(y) f(y, \omega, T) y^k \right]^\star}_{ = C C^\star = |C|^2 \ge 0}\\
            & \ge 0.
        \end{align}
    \end{widetext}


\bibliography{BibtexDatabase}

\end{document}